\documentclass[prl,twocolumn,nofootinbib,preprintnumbers,amssymb,amsfonts,amsmath,superscriptaddress,showpacs,hyperref]{revtex4-1}

\usepackage{graphicx}
\usepackage{bm}
\usepackage{color}
\usepackage{url}
\usepackage{epstopdf}

\newcommand{\eV}{{\rm eV}}

\newcommand{\GeV}{{\rm GeV}}
\newcommand{\TeV}{{\rm TeV}}

\newcommand{\Mpc}{{\rm Mpc}}
\newcommand{\kpc}{{\rm kpc}}

\newcommand{\km}{{\rm km}}

\newcommand{\s}{{\rm s}}

\newcommand{\Mpl}{M_{\rm Pl}}

\begin{document}

\title{Astrophysical constraints on Planck scale dissipative phenomena}

\author{Stefano Liberati}
\affiliation{SISSA, Via Bonomea 265, 34136, Trieste, Italy {\rm and} INFN, Sezione di Trieste}

\author{Luca Maccione}
\affiliation{Arnold Sommerfeld Center, Ludwig-Maximilians-Universit\"at, Theresienstra{\ss}e 37, 80333 M\"unchen, Germany} 
\affiliation{Max-Planck-Institut f\"ur Physik (Werner-Heisenberg-Institut), F\"ohringer Ring 6, 80805 M\"unchen, Germany} 

%
\begin{abstract}
The emergence of a classical spacetime from any quantum gravity model is still a subtle and only partially understood issue. If indeed spacetime is arising as some sort of large scale condensate of more fundamental objects then it is natural to expect that matter, being a collective excitations of the spacetime constituents, will present modified kinematics at sufficiently high energies. We consider here the phenomenology of the dissipative effects necessarily arising in such a picture. 
Adopting dissipative hydrodynamics as a general framework for the description of the energy exchange between collective excitations and the spacetime fundamental degrees of freedom, we discuss how rates of energy loss for elementary particles can be derived from dispersion relations and used to provide strong constraints on the base of current astrophysical observations of high energy particles. 
\end{abstract}
\preprint{LMU-ASC 67/13; MPP-2013-272}
\maketitle

{\em Introduction.---} Quantum Gravity (QG) has been for many years now a frustrating effort, a theoretical research program devoid of a strong guidance from experiments and observations.   The so called QG phenomenology field has been a recent answer to this frustration, a quantum leap to bypass the detailed technical issues about the functionality of QG models and to actually focus on the possible scenarios for the emergence of a classical spacetime from these discrete quantum models. 

Such an issue is at the core of many, if not all, the most pressing questions of the viable models of QG 
as the microscopic theory does not necessarily share all the symmetries of a classical spacetime. How and by which mechanisms such symmetries can be recovered is obviously of the uttermost importance in order to subject these models to the observation test.

In this view it has often been conjectured (and in some QG toy models observed) that Lorentz symmetry might be violated in QG (see e.g.~\cite{Liberati:2013xla} for a recent review), and indeed much attention has been given to constraints on modified dispersion relations associated to the breakdown of rotation and boost invariance~\cite{Mattingly:2005re}. However, much less attention has been given to dissipative phenomena induced by the possible exchange of energy between matter and some fundamental constituents that could be at the base of the emergent spacetime.

Generically, for the propagation of perturbations in a medium in a causality preserving theory, dispersion and dissipation are related by the so called Kramers--Kronig (KK) relations~\cite{Parentani:2007uq},
according to which dispersive effects can only arise if also dissipative effects are present.   Therefore, in an emergent gravity picture considering dispersive effects neglecting dissipative ones seems inconsistent. 

Possible dissipative effects in the context of fundamental theories of gravity have been discussed, e.g. in \cite{Dvali:2000rx}, as an infrared signal of the coupling of matter fields with gauge modes propagating in extra-dimensions, while dispersive phenomena in emergent spacetime have been discussed in causal set theory \cite{Dowker:2004,Philpott:2009}. Effective field theories (EFT) with broken Lorentz invariance (LI) in the ultraviolet sector and showing both dispersion and dissipation have been studied in \cite{Parentani:2007uq} where also the links with phenomenological QG and some brane world scenarios have been discussed. 
According to \cite{Parentani:2007uq} it is not justified to assume, as it has been done in the past, that modified dispersion relations arising in QG models do not contain dissipative terms, if LI breaking is to be a dynamical process. The presence of dissipation can possibly invalidate previous limits. 

In this Letter we show for the first time limits on dissipative effects derived from high-energy astrophysics observations. In order to be as generic as possible, we do not adopt directly the EFT description given in \cite{Parentani:2007uq} ---  which would require a detailed modelling --- but we shall rather appeal to the so called Analogue gravity framework \cite{Barcelo:2005fc} describing the dynamics of the matter propagating on an emergent spacetime as collective excitation in hydrodynamics. 

We remark that there are systems whose particular internal symmetries prevent dissipative effects from being present at the lowest order in the hydrodynamics approximation, without violating the KK relations. Bose--Einstein condensates are well-known analogue gravity examples of this fact. In this sense, the here presented strong observational bounds imply that any emergent spacetime scenario should behave similarly, i.e.~its hydrodynamic description should be close to that of a superfluid.


{\em An Analogue Gravity lead.---} The framework of Analogue gravity \cite{Barcelo:2005fc}, has been widely used as a test field for the phenomenology of quantum field theory on a curved spacetime. Linear perturbations in an inviscous and irrotational flow propagate as fields on a curved spacetime, whose metric, the so called acoustic metric, depends on the background flow density and velocity \cite{Barcelo:2005fc}. In the presence of kinematical viscosity such a picture is changed however. In this brief section we shall review the simple case of an irrotational, barotropic fluid with non-zero kinematic/shear viscosity, $\nu$, and show how this can naturally provide a modified dispersion relation which entails dissipative effects (we consider here for simplicity an incompressible fluid i.e.~with zero bulk viscosity). A detailed discussion can be found in \cite{Visser:1997ux}.  Technically, the collective excitations of the medium are gravitons, hence we should apply our reasoning only to these particles (albeit one expects that also matter should emerge in complete QG scenarios). Nonetheless, radiative corrections-induced percolation of Lorentz breaking terms (from the gravitational to the matter sector) is generically foreseeable (see e.g.~\cite{Pospelov:2010mp} for the case of dispersive effects).

If the background fluid flow is at rest and homogeneous (bulk velocity $\vec v_0 = 0$, with position independent density $\rho_0$ and speed of sound $c$) then the
viscous wave equation for the perturbations in the velocity potential, $v^\mu=\nabla^\mu \psi$ is simply
\begin{equation}
\partial_t^2  \psi_1  = c^2 \nabla^2 \psi_1  + {4\over3} \nu \; \partial_t \nabla^2 \psi_1\, ,
\label{E-lamb}
\end{equation}
where $\psi_1$ is a linear perturbation of $\psi$.
This equation may be found, e.g., in \cite{lamb:1916hydro} and is explicitly derived in \cite{Visser:1997ux}.

In order to find the corresponding dispersion relation one can adopt as usual the so-called Eikonal approximation in the form 
%
$\psi_1 = a(x) \exp(-i[\omega t - \vec k\cdot \vec x]\,)$,
%
with $a(x)$ a slowly-varying function of position. 
Then the viscous wave equation in the Eikonal approximation reduces to
$-\omega^2 + c^2 k^2  - i\nu \, {4/3} \omega k^2 = 0$
%
%
yielding the following dispersion relation for sound waves
\begin{equation}
\omega =  \pm  \sqrt{c^2 k^2 - \left({2\nu k^2\over3}\right)^2 }  - i {2\nu k^2\over3}.
\end{equation}
The first term specifically introduces dispersion due to viscosity,
while the second term is specifically dissipative. The previous equation can be further simplified to
\begin{equation}
\omega^2 \simeq  c^2 k^{2} \left[1 - i \frac{4}{3}\frac{\nu k}{c}-\frac{8}{9} \left(\frac{\nu k}{c}\right)^{2}+i \frac{8}{27} \left(\frac{\nu k}{c}\right)^{3}\right]\;,
\label{eq:viscosity}
\end{equation}
up to higher orders of $(\nu k/c)$. 
This is a concrete example of how modified dispersion relations due to the underlying microscopic structure of an emergent spacetime can be endowed with dissipative terms. In this case the lowest order dissipative term, which is ruled by the same microscopic scale provided by the viscosity, would appear {\em at lower energies} than the dispersive, quartic term.

Such dissipative dispersion relations clearly violate unitarity. However in this toy model dissipation is due to energy exchange with extra degrees of freedom which, being not observed, are traced away \cite{Parentani:2007uq}. In this sense, (apparent) dissipation can be a signal of extra-degrees of freedom in putative  (unitarity preserving) fundamental theories being neglected in the effective theory.


{\em Generalized dissipative hydrodynamics.---} If we now come back to the problem of the phenomenology associated to an emergent spacetime from some, unspecified, QG model, we are faced with a set of pressing questions, which basically deal with our ignorance of the models and the viable mechanisms leading to a classical spacetime. 

In this sense one quite general approach might consist in assuming that at sufficiently low scales any QG theory will allow to describe the propagation of matter (or gravitons) on the emergent spacetime along the equations one could derive from hydrodynamics. Implicitly we are assuming that a description of matter as collective excitations above the spacetime medium is possible at scales much longer than the typical scales of the fundamental constituents interactions. This is tantamount to assume that some EFT description is viable given that hydrodynamics, even dissipative one, can be described within this formalism \cite{Endlich:2012vt}.

When adopting hydrodynamics as a large scale model of an emergent spacetime, it is quite interesting to keep in mind that the above discussed dissipation appears in a gradient expansion as a first order correction to the perfect fluid equations.  
In general, higher order terms can be considered as well, and such operators will show a similar structure to the last term on the right hand side of Eq.~\eqref{E-lamb}, i.e.~they will be generically of the form~$\partial_t \nabla^{n}$. Hence, dissipative terms will always appear in the dispersion relation with odd powers of the three momentum $k$ once at high energy one takes $E\approx k$. 

The generalised  Navier-Stokes equation will then read
\begin{equation}
\partial_t^2  \psi_1  = c^2 \nabla^2 \psi_1  + \sum_{n=2}^{\infty}{4\over3} \nu_{n} \; \partial_t \nabla^{n} \psi_1\, ,
\label{E-lamb-full}
\end{equation}
leading to the following dispersion relation
\begin{eqnarray}
\nonumber & &\omega^{2} = c^{2}k^{2}-i\frac{4}{3}\nu_{2}ck^{3}\\ 
\nonumber & &+ \sum_{j=1}^{\infty}(-)^{j+1}\mathcal{L}_{6j-3}^{3j-1}k^{6j-2}+ i\sum_{j=1}^{\infty}(-)^{j+1}B_{2j}^{6j-2}k^{6j-1} \\
\nonumber & &+ \sum_{j=1}^{\infty}(-)^{j}\mathcal{L}_{6j-1}^{3j}k^{6j}+ i\sum_{j=1}^{\infty}(-)^{j}A_{6j}k^{6j+1} \\
\nonumber & &+ \sum_{j=1}^{\infty}(-)^{j+1}\mathcal{M}_{6j+1,2j+1}^{3j+1}k^{6j+2} + i\sum_{j=1}^{\infty}(-)^{j+1}A_{6j+2}k^{6j+3} \;,
\end{eqnarray}
with $A_{i}=4/3\nu_{i}c$, $B_{i}^{j}=8/27\nu_{i}/c+4/3\nu_{j}c$, $\mathcal{L}_{i}^{j}=4/3\nu_{i}c-8/9\nu_{j}^{2}$ and $\mathcal{M}_{i,j}^{k}=4/3\nu_{i}c+8/27\nu_{j}^{3}/c-8/9\nu_{k}^{2}$. 

The structure of the modified dispersion relation is such that it is not possible to proceed simply order-by-order to account for all the possible terms. Actually, one should not attempt to identify the index of the expansion $j$ with the power index of the derivative term $\nabla^n$ in (\ref{E-lamb-full}). The contribution of each term of order $n$ in the derivative expansion is instead to be searched for in all those terms which at some order $j$ will contain the parameter $\nu_n$.

As expected, Eq.~(\ref{E-lamb-full}) shows alternating dissipative and dispersive terms with odd and even powers of $k$ respectively. Assuming that the origin of these deviations from the perfect fluid limit is related to the behavior of the ``spacetime fluid" close to the Planck scale, it is natural to rescale the coefficients of Eq.~(\ref{E-lamb-full}) by suitable powers of the Planck energy so to make the coefficient dimensionless and make explicit the suppression of higher powers terms (assuming, as a matter of naturalness, that the remaining dimensionless coefficients are a priori roughly of the same magnitude). 

Let us start truncating the above dispersion relation to the lowest order, $n=2$, so regaining (\ref{eq:viscosity}), with a suitably rescaled coefficient as described above. We get
\begin{equation}
\omega^2 = c^2 k^2 - i \sigma_2 c^2 \frac{k^3}{\Mpl}\, ,
\label{test-mod}
\end{equation}
where $\sigma_2=(4\nu_2 \Mpl)/3c$ is the dimensionless coefficient controlling the magnitude of the Lorentz violation (LV) and $\Mpl=1.22\times10^{19}~\GeV$. 

We have neglected extra --- non-derivative expansion generated --- {\em dispersive} effects at the same $k^{3}$ order 
(e.g.~the CPT odd dimension five operators characterizing the EFT photon dispersion relation~\cite{Myers:2003fd}), however they are not relevant in this context. Indeed, taking the best constraint $\xi \lesssim 10^{-16}$ on such operators~\cite{Gotz:2013dwa}, 
the maximal correction to the photon energy in the ultra relativistic limit is $\omega/k-1 \simeq \xi/2 k/\Mpl \sim 4\times10^{-31}$ at 100 TeV. On the other hand, the constraints we shall place on $\sigma$ are so strong so to not invalidate the aforementioned constraints on modified dispersion.
%


{\em Computing the rate.---} A major obstruction for casting an observational constraint on Eq.~(\ref{test-mod}) consists in the fact that dissipative effects would imply to work with a non-unitary EFT. This could be avoided by resorting to a system-environment Ansatz \cite{Caldeira:1981rx,Parentani:2007uq} but in this case we would work with a more complicate system for which the results can be even model dependent. We will instead follow the lead of \cite{Parentani:2007uq} for obtaining a generic estimate of the energy loss rate.

Dispersive effects in vacuum are fully specified by the imaginary part of the self-energy. The energy loss rate $\Gamma$ can be readily obtained by inverse Fourier transforming the retarding Green function corresponding to the dispersion relation of Eq.~(\ref{test-mod}). In our case, $\Gamma$ can then be written as
\begin{equation}
\sigma_2 c^2 \frac{k^3}{\Mpl} \equiv 2 \omega \Gamma \;,
\end{equation}
where $\Gamma$ represents the energy loss rate in the underdamped regime $\Gamma \ll k$. 
Assuming $\omega \sim k$ for our purposes, we can identify the energy loss rate 
%
$\Gamma\approx \sigma_2 k^2/(2\Mpl)$.
%
This result agrees with what one would find by applying na\"ively the well known relations for unstable particles in the Breit--Wigner formalism.
Being the lifetime $\tau = \hbar/\Gamma$, 
we have now all the necessary information to cast our constraint.


{\em Constraints on ``spacetime viscosity".---} 
For an ultra-relativistic particle with momentum $p$ traveling over a long distance $D$, a constraint is obtained by requiring its lifetime $\tau$ to be larger than the propagation time $D/c$, that is $\tau\geq D/c$ or ${c\hbar/ \Gamma} \geq D$.
%
%

\begin{figure}[htbp]
\begin{center}
\includegraphics[width=0.30\textwidth]{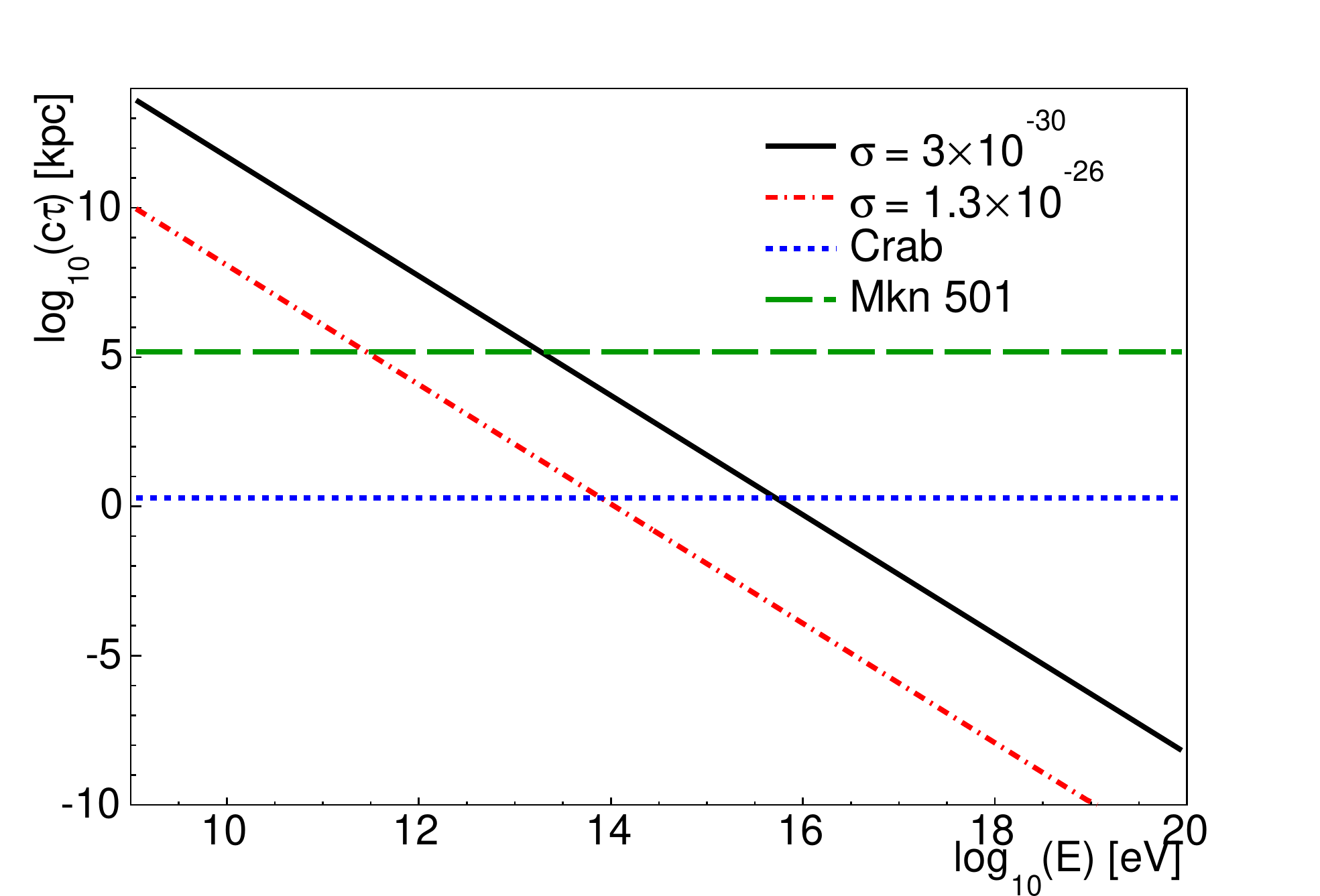}
\caption{Mean free path of photons subject to dissipation versus energy. The blue, dotted horizontal line represents the distance of the Crab nebula, while the green long-dashed horizontal line is for the reference distance of Mkn 501.}
\label{fig:LifeTime}
\end{center}
\end{figure}

Let us consider the observed 80 TeV photons from the Crab nebula (see \cite{Meyer:2010tta} for an up-to-date compilation of spectral data) which is at a distance $D_{\rm Crab} \simeq 1.9~\kpc$. We then obtain (see also Fig.~\ref{fig:LifeTime})
\begin{equation}
\sigma_2 \leq {2c\hbar \over D_{\rm Crab}(80~\TeV)^2}\Mpl \approx 1.3\times10^{-26}\;,
\end{equation}
%
%
Noticeably, in the standard model of the Crab nebula, such very high-energy photons are produced by inverse Compton scattering of electrons and positrons accelerated and propagating in the nebula. Therefore, the same constraint can be applied for such leptons, by assuming, conservatively, that they have at least the same energy as the produced photons (by energy conservation), and by properly rescaling the propagated distance to a parsec, which is the typical size of the nebula. Hence, the constraint is weakened by a factor $10^3$ for electrons/positrons. 

A constraint of order $2\times10^{-27}$ can be obtained for neutrinos, given the detection of a bunch of extraterrestrial neutrinos with energies between $30$ and $250~\TeV$ by IceCube \cite{IceCube1,IceCube2,IceCube3}. Their energy spectrum is consistent with a single power-law \cite{Anchordoqui:2013qsi}. Assuming conservatively that they are of galactic origin, we can set their propagation distance $D\simeq8~\kpc$ (this is approximately the distance between the Earth and the galactic center). Taking then for definiteness $E_{\nu}\simeq 100~\TeV$, the constraint for neutrinos would be about 6 times better than the one we placed for photons.

Even stronger constraints can be placed if extra-galactic objects are considered. For example, the Mkn 501 has been observed up to 24~TeV \cite{Aharonian:1999vy}. Its redshift is estimated as $z=0.034$, which corresponds to an effective distance\footnote{The effective distance is computed as $D = 1/H_{0}\int_{0}^{z}dz' (1+z')^{2}/\sqrt{\Omega_{\Lambda} + \Omega_{m}(1+z')^{3}}$, and
we used the values of $H_{0} = 71~\km/\s/\Mpc$, $\Omega_{\Lambda}=0.7$ and $\Omega_{m}=0.3$.} 
of $\sim147~\Mpc$. The implied constraint on $\sigma_2$ is then of the order of $3\times10^{-30}$ (see again Fig.~\ref{fig:LifeTime}).

One possible caveat can arise from the fact that we do not know observationally the energy spectrum of the photons leaving these sources. This might be of particular relevance for variable objects, like Mkn 501. It may therefore happen that the initial spectrum and the dissipation effects combine to yield by chance the observed spectrum at Earth. However, the energy loss rate $\Gamma$ is strongly dependent upon energy. The observed very high-energy spectrum of the Crab nebula, on the other hand, is remarkably featureless up to the highest energies, thereby hinting at the absence of energy dependent effects.

Even stronger constraints can in principle be derived considering the extragalactic propagation of ultra-high energy cosmic rays, with energy above $10^{18}~\eV$. However, in that case we should first understand how dissipation would affect compound particles, which goes much beyond the scope of this Letter. 
 Finally, also gravitational waves could in principle provide constraints in case of detection. Unfortunately, current experiments are sensitive to waves which are far too low energy (below 1~Hz) for providing meaningful constraints. 


{\em Constraints on higher-order terms.---} The constraint derived above, places a very strong limit on the coefficient $\nu_{2}$ in the expansion of Eq.~(\ref{E-lamb-full}). However, higher order terms are not naturally as suppressed given that e.g.~$\nu_3$ and $\nu_4$, being of different dimensionality of $\nu_2$, do not need naturally to be of the same magnitude. Indeed, nothing prevents a medium with effectively zero kinematic viscosity to have non zero higher order transport coefficients, otherwise we would end up with the unphysical case of a fluid being perfect at all scales.  

We consider then such higher order terms in our expansion Eq.~(\ref{E-lamb-full}) and in particular focus on the next term producing a dissipative contribution to the dispersion relation, assuming $\nu_2\approx 0$. Also in this case one can check that given the best constraints on the $\sigma_3\equiv (4\nu_3 \Mpl^2)/3c$ coefficient of the dispersive term of order $O(k^4)$, $\sigma_3\approx O(10^{-7}\div10^{-8})$ see e.g.~\cite{Liberati:2013xla}, we can safely neglect its effect on the energy of photons at our reference energy of about 100 TeV.

The next contribution is then imaginary $\omega^2 \simeq k^{2}+i2/3\nu_{4}/ck^{5}$. 
Note however that this has opposite sign with respect to the one induced by the $\nu_{2}$ term. Assuming that $\nu_{4}>0$, this implies that the effect induced at this order is not dissipation, rather amplification, with the matter field increasing its intensity as it propagates. 

Albeit a full EFT derivation of such higher order derivative terms (e.g.~along the approach of \cite{Endlich:2012vt}) would be required to fully clarify this issue, we consider highly implausible that such stimulated growth of excitations would be allowed by the second law of thermodynamics. In fact, in an hydrodynamical interpretation the proliferation of high energy collective excitations of the substratum would correspond to a net decrease of the total entropy of the system.  We hence expect that in a full fledged EFT derivation of the hydrodynamics, 
the coefficients ruling dissipative effects will always come with appropriate signs so to avoid amplification.

Fortunately, the choice of the sign of the coefficient $\nu_4$ is not very relevant for casting a constraint\footnote{ Note that while in principle a cancellation of the effects induced by the $\nu_{2}$ and a negative $\nu_{4}$ could accidentally happen at some energy, their different energy dependence implies that the two contributions cannot erase each other at all energies.}, as in any case fluxes from well known astrophysical objects such as the Crab nebula would be unacceptably modified by either energy loss or gain.
Similarly to what we did in Eq.~(\ref{test-mod}), we can study a dispersion relation
%
$\omega^2 = c^2 k^2 \pm i |\sigma_{4}| c^2 k^5/\Mpl^{3}\;,$ where $\sigma_4\equiv(4\nu_4 \Mpl^3)/3c$,
%
leading to a energy loss/gain rate $\Gamma_{4} = |\sigma_{4}|/2\cdot k^{4}/\Mpl^{3}$. Again, the absence of energy-dependent energy loss/gain effects during the propagation of very high-energy photons from the Crab Nebula implies a limit $\sigma_{4}\lesssim300$, while considering the spectrum of Mkn 501 yields a constraint $\sigma_{4}\lesssim 0.5$. Hence, already at the next order of dissipation the strength of the constraints is greatly reduced. 


{\em Conclusions.---} While dispersive Lorentz breaking effects have been widely studied in the past, basically no constraints have been cast so far on departures from exact Lorentz invariance due to high-energy dissipative effects. 
 In this Letter, we carried out the first systematic discussion of such dissipative terms adopting hydrodynamics as a very general framework within which the emergence of a classical spacetime below the Planck energy could be described.  The bounds we obtained at the lowest order (what one might call the spacetime viscosity) are indeed extremely tight, pushing the scale for such dissipative effects well beyond the Planck scale by several orders of magnitude. 

Unfortunately it is not possible, missing a detailed microscopic understanding about the origin of the dissipative dispersion relation \eqref{test-mod}, to link this constraint to some physical property of the underlying theory.  Once quantum gravity scenarios will be able to fully describe the emergence of spacetime (and of matter), the bounds on the so derived hydrodynamic coefficients will tell us more about the theory. Nonetheless, the very tight constraints here obtained are already providing the very important information that any viable emergent spacetime scenario should provide a hydrodynamical description of the spacetime close to that of a superfluid.

%

Finally it is worth stressing that higher order dissipative terms can and in principle should be considered. For example nothing forbids such terms in superfluids (which have zero viscosity) to be non-zero. Similarly, if some fundamental, custodial, symmetry of the underlying, quantum gravitational system would forbid the above mentioned ``spacetime viscosity" term still one could expect non-zero dispersive $O(k^4)$ and dissipative $O(k^5)$ terms to appear. These are sufficiently high energy modifications for which we do have relatively weak constraints on dispersion and basically, as shown above, no constraints on dissipation. We think that such dispersive-dissipative relations deserve further exploration and
we hope that the study presented here can be of some stimulus for such further investigations in a near future.

{\em Acknowledgments:} We thank Andrea Gambassi for illuminating discussions and David Mattingly for reading a preliminary draft of this Letter and providing useful insights. We also thank Stefano Finazzi, Ted Jacobson, Arif Mohd and Matt Visser for useful remarks.


\begin{thebibliography}{999}

\bibitem{Liberati:2013xla} 
  S.~Liberati,
  Class.\ Quant.\ Grav.\  {\bf 30}, 133001 (2013)
  [arXiv:1304.5795 [gr-qc]].
  
 \bibitem{Mattingly:2005re} 
  D.~Mattingly,
  Living Rev.\ Rel.\  {\bf 8}, 5 (2005)
  [gr-qc/0502097].
  
        \bibitem{Parentani:2007uq}
  R.~Parentani,
  PoS QG {\bf -PH} (2007) 031
  [arXiv:0709.3943 [hep-th]].


  \bibitem{Dvali:2000rx}
  G.~R.~Dvali, G.~Gabadadze and M.~A.~Shifman,
  Phys.\ Lett.\ B {\bf 497} (2001) 271
  [hep-th/0010071].
  


  \bibitem{Dowker:2004}
  F.~Dowker, J.~Henson and R.~D.~Sorkin,
  Mod.\ Phys.\ Lett.\ A {\bf 19}, 1829 (2004)

  \bibitem{Philpott:2009}
 L.~Philpott, F.~Dowker and R.~D.~Sorkin,
  Phys.\ Rev.\ D {\bf 79}, 124047 (2009)
  
  
\bibitem{Barcelo:2005fc} 
  C.~Barcelo, S.~Liberati and M.~Visser,
  Living Rev.\ Rel.\  {\bf 8}, 12 (2005)
  [Living Rev.\ Rel.\  {\bf 14}, 3 (2011)]
  [gr-qc/0505065].
  
\bibitem{Visser:1997ux} 
  M.~Visser,
  Class.\ Quant.\ Grav.\  {\bf 15}, 1767 (1998)
  [gr-qc/9712010].
  
  \bibitem{Pospelov:2010mp}
  M.~Pospelov and Y.~Shang,
  Phys.\ Rev.\ D {\bf 85} (2012) 105001
  [arXiv:1010.5249 [hep-th]].
  
\bibitem{lamb:1916hydro}
H.~Lamb,
  ``Hydrodynamics'',
University Press, 1916.

\bibitem{Endlich:2012vt} 
  S.~Endlich, A.~Nicolis, R.~A.~Porto and J.~Wang,
  arXiv:1211.6461 [hep-th].
  


  \bibitem{Myers:2003fd} 
  R.~C.~Myers and M.~Pospelov,
  Phys.\ Rev.\ Lett.\  {\bf 90}, 211601 (2003)
  [hep-ph/0301124].
  
  \bibitem{Gotz:2013dwa} 
  D.~Gotz, S.~Covino, A.~Fernandez-Soto, P.~Laurent and Z .Bosnjak,
  arXiv:1303.4186 [astro-ph.HE].
  
  \bibitem{Caldeira:1981rx} 
  A.~O.~Caldeira and A.~J.~Leggett,
  Phys.\ Rev.\ Lett.\  {\bf 46}, 211 (1981).
  

%
%
\bibitem{Meyer:2010tta}
  M.~Meyer, D.~Horns and H.~-S.~Zechlin,
  Astronomy and Astrophysics {\bf } (2010) , Volume 523, id.A2
  [arXiv:1008.4524 [astro-ph.HE]].

\bibitem{IceCube1}
C. Kopper [for the IceCube Collabotration], 
{\it ``Observation of PeV neutrinos in IceCube''}, talk given at the IceCube Particle Astrophysics Symposium (IPA-2013), Madison, Wisconsin, 13-15 May 2013. \url{http://wipac.wisc.edu/meetings/home/IPA2013}

\bibitem{IceCube2}
N. Kurahashi-Neilson [for the IceCube Collaboration],
{\it ``Spatial Clustering Analysis of the Very High Energy Neutrinos in IceCube''}, talk given at the IceCube Particle Astrophysics Symposium (IPA-2013), Madison, Wisconsin, 13-15 May 2013.

\bibitem{IceCube3} 
N. Whitehorn [for the IceCube Collabotration], 
{\it ``Results from IceCube''}, 
talk given at the IceCube Particle Astrophysics Symposium (IPA-2013), Madison, Wisconsin, 13- 15 May 2013.

\bibitem{Anchordoqui:2013qsi}
  L.~A.~Anchordoqui, H.~Goldberg, M.~H.~Lynch, A.~V.~Olinto, T.~C.~Paul and T.~J.~Weiler,
  arXiv:1306.5021 [astro-ph.HE].
  
  \bibitem{Aharonian:1999vy}
  F.~Aharonian [HEGRA Collaboration],
  Astron.\ Astrophys.\  {\bf 349} (1999) 11
  [astro-ph/9903386].
  
  
\end{thebibliography}
\end{document}